\documentclass[aps,preprint,showpacs,showkeys,superscriptaddress]{revtex4-1}

\usepackage{dcolumn}
\usepackage{bm}
\usepackage{amssymb}
\usepackage{amsfonts}
\usepackage{amsmath}
\usepackage{graphicx}
\usepackage{epstopdf}
\usepackage{float}
\usepackage{relsize}
\usepackage{mathtools}
\usepackage{booktabs}
\usepackage{placeins}
\usepackage{array}
\begin{document}

\title{Black holes as gravitational mirrors}

\author{L. C. N. Santos}
\email{luis.santos@ufsc.br}

\affiliation{Departamento de F\'isica, CFM - Universidade Federal de Santa Catarina; C.P. 476, CEP 88.040-900, Florian\'opolis, SC, Brasil.} 

\author{F. M. da Silva}
\email{franmdasilva@gmail.com}

\affiliation{Departamento de F\'isica, CFM - Universidade Federal de Santa Catarina; C.P. 476, CEP 88.040-900, Florian\'opolis, SC, Brasil.}

\author{C. R. Muniz}
\email{celio.muniz@uece.br}

\affiliation{Departamento de Física, FECLI, Universidade Estadual do Ceará, Av. Dário Rabelo, Iguatu 63500-518, Brazil}

\author{V. B. Bezerra}
\email{valdir@fisica.ufpb.br}

\affiliation{Departamento de Física, CCEN--Universidade Federal da Paraíba; C.P. 5008, CEP  58.051-970, João Pessoa, PB, Brazil}

\begin{abstract}
Retrolensing is a gravitational lensing effect in which light emitted by a background source is deflected by a black hole and redirected toward the observer after undergoing nearly complete loops around the black hole. In this context, we explore the possibility of seeing objects of the solar system in past eras through telescope observations by using black holes as a gravitational mirror. We consider the motion of the light around Reissner–Nordström space–time and discuss the properties of the trajectories of boomerang photons. It was shown that, depending on the angle of emission and the position of the source, the photons could return to the emission point. 
Afterward, we explore the possibility of considering the returning photons in retrolensing geometry where the observer is between the source and the lens in which two classes of  black holes are explored: The~supermassive Sgr A* black hole at the galactic center and a nearby stellar black hole. For~the first time in the literature, we propose the study of the returning photons of planets instead of stars in retrolensing geometry.

\end{abstract}


\maketitle

\preprint{}

\volumeyear{} \volumenumber{} \issuenumber{} \eid{identifier} \startpage{1} %
\endpage{}
\section{Introduction}
By observing the light of distant galaxies, we are seeing them as they were when the {light}
 left them { millions or even billions of} years ago. This provides the possibility that the light of these astronomical objects carries information about their structure in different eras of space–time. Spectral lines can be used to determine the value of physical quantities such as the angular speed and mass of the galaxy, and so on. The idea of looking back in time by observing the light emitted in the past deals with rays of light that travel {from some minutes, such as the light that reaches us from the Sun to even billions of light-years, in the case of light coming from the most distant galaxies}. The practical utility of this type of observation can be viewed{, for example,} in the detection of the light of distant galaxies that allows us to see {what the universe was like} billions of years ago.

On the other hand, general relativity (GR) tells us that the light emitted from these objects actually propagates in a curved path due to the distribution of energy and matter in the universe. Indeed, one of the first observational triumphs of GR was the correct description of the bending of light in the vicinity of the Sun with the weak-field consideration \cite{retro1}. 
Since then, the study of the gravitational lensing phenomena has been revisited several \mbox{times \cite{retro2,retro3,retro4,retro5,retro6,retro7,retro8}.} Compact objects such as black holes have a gravitational field so strong that light rays from the stars and galaxies can orbit the black hole several times before reaching a distant observer in space–time \cite{page,muller,cramer}. In this context, the strong gravitational regime imposes intense effects on the structure of space–time, and the geodesic equations for the bending of light in this regime must be solved without a weak field approximation. Recently, the Event Horizon Telescope (EHT) collaboration made the first image of the M87$^*$ black hole, observing an intricate pattern of the paths of light around it \cite{retro11,retro12,retro13}.

In this way, as in the case of the orbit of massive particles, photons can also circulate around a black hole \cite{retro9,retro10}. For example, the region where {$r=3~GM/c^2$} 
 called the photon sphere, is the lower bound for stable orbits for the uncharged Schwarzschild black hole.  At~this place, the photon may either escape to infinity or fall into the black hole. Furthermore, there are additional regions in space–time where photons can be emitted  and loop around the black hole, returning to the emission point depending on the emission angle of the photons and the distance of the black hole. Note that there is no length limit for the radius of the orbit for these boomerang photons; consequently, a ray of light from the Earth to the compact radio source Sgr A* can loop around the black hole and return to the solar system. In this case, when the photons return, they carry information about the rays of light emitted by a source 52,000 years ago.

A flash of light that returns to the emission point can also be used to determine the mass and the distance of a black hole through the apex angles associated with orbits obtained from the geodesic
equation in Schwarzschild space–time \cite{muller}. A pedagogical discussion on the uncharged Schwarzschild black hole can also be found in \cite{stuckey}. This work discusses the trajectories of photons that circle around black holes. In \cite{cramer}, the author extends the use of the static black hole as a gravitational mirror to the case of an uncharged Kerr black hole, giving a precise description of the concept of boomerang photons.  

Due to the fact that compact objects can deflect the light ray paths to large bending angles, a particular type of gravitational lensing has been explored recently. Holz and Wheeler introduced the concept of retrolensing, where the observer may be between the source and the lens \cite{retro14}. In this work, the authors estimate the apparent magnitude of retrolensing events by considering black holes as lenses placed at the border of our solar system. The Sgr A* black hole at the galactic center as a retrolens for a star at a close distance was considered in \cite{retro15}. In this case, the results provide evidence that such an event could have high magnitude as a consequence of the vicinity of the star to the Sgr A* black hole. Other advances in retrolensing systems have been made in the \mbox{following articles \cite{retro16,retro17,retro18,retro19,retro20,retro21,retro22}.} {\mbox{In Ref. \cite{naoki1},} whose approach we closely follow, retrolensing is analyzed for the charged Reissner–Nordström black holes, while in Ref. \cite{retro20}, the electric charge is replaced by the Weyl tidal charge, in the form $Q^2\to -W^2$, which acts in consonance with the black hole mass, in a braneworld model.} Thus, although retrolensing has been extensively studied in theoretical and numerical models, it remains an open problem in \mbox{observational astrophysics. }
 
The aim of this article is to present a study of trajectories of returning photons in a retrolensing geometry considering bodies in the solar system. As {a particular} system, Earth as a source in the retrolensing geometry is considered when we propose an apparatus to improve the magnification of the image. In this way, our focus is on the properties of the image of the source of photons in addition to the existing studies of retrolensing, where the goal is to determine the properties of the path of light around a source of the strong gravitational field.   
To this end,  we structure the paper as follows: In {Section \ref{sec2},} we review the Reissner–Nordström space–time and discuss some basic aspects concerning the trajectories of particles in this geometry. In {Section \ref{sec3},} we study photon trajectories in a general static spherically symmetric space–time, and in {Section \ref{sec4},} we explore the trajectories of the photons that return to their emitter and find out the conditions for this type of motion. In {Section \ref{sec5},} we propose a scheme to address the lens geometry associated with the Earth as a source and an orbiting satellite as an observer in a retrolensing event. Finally, in {Section \ref{sec6}}, we discuss our results.

\section{Spherically Symmetric Space–Time}\label{sec2}

It is possible to consider the trajectories of photons in space–time with a general spherical form. For this purpose, the line element that represents this symmetry can be written in the form
\begin{equation}
    ds^2 = -B(r)dt^2 + A(r)dr^2 + r^{2}d\theta^2 + r^{2}\sin{\theta}^{2}d\varphi^2,
\label{eq1}
\end{equation}
where $-\infty < t <+ \infty,0 \leq r < + \infty,0 \leq \theta \leq 2\pi$ and $0 \leq \varphi \leq \pi$. 

\subsection*{Charged Black {Hole}}

In solving Einstein's equation, the functions $A(r)$ and $B(r)$ can be written explicitly. In~the case of the Reissner–Nordström space–time manifold,  these functions are then 
\begin{equation}
    B(r)=A(r)^{-1} = 1-\frac{2M}{r} + \frac{Q^{2}}{r^{2}},
    \label{eq2}
\end{equation}
with $Q$ being the black hole charge. In this line element, the constant parameter $M$ is the mass of the stellar object. In the case of a black hole of mass $M$, the trajectories of the photons are so strong that light can even return to the point of emittance; this point will be discussed in detail in the following sections. In the limit where $Q$ goes to zero, the space–time metric becomes the Schwarzschild metric, and the equivalent of the condition {$r=2~M$} 
 in the charged case is given by 

\begin{align}
      r_{\pm}=M\pm\sqrt{ M^2 -  Q^2}. 
    \label{eq3}
\end{align}
 In the case that $M^2 > Q^2$, the black hole has two coordinate singularities, $r_{+}$ and $r_{-}$. Thus, the radial coordinate $r$ is regular in the regions $r_{+}<r<\infty$, $r_{-}<r<r_{+}$ and $0<r<r_{-}$. In contrast, if $M^2 < Q^2$, the space–time is non-singular for all values of $r>0$; however, the singularity at $r=0$ remains. When the black hole mass satisfies {the relation $ M^2 = Q^2$,} there is only one horizon; this type of charged black hole is extremal. We are interested in the behavior of light rays in the region $r>r_{+}$, i.e., outside the horizon. The proper time $\tau$ elapsed for a static observer is related to the time coordinate $t$ as $\Delta\tau = \sqrt{1-2~M/r}\Delta t$ with $Q=0, $ which means that time passes more slowly near the black hole.

\section{Photon Trajectories in a General Static Spherically Symmetric
Space–Time}\label{sec3}

{An important physical quantity to the study of photon trajectories is the point where the potential associated with the photon trajectory $V(r)$ reaches an extremum $r_m$ (the derivation of $r_m$ can be found in Appendix \ref{appenA}). }
As can be seen in Figure \ref{fig1}, $r_m$ corresponds to the extremum values of the curves of the effective potential as a function of $r$ (in terms of $M$). The solid line denotes {the} potential of a photon in the space–time of a non-charged black hole with $r_m=3~M$. In contrast, the dashed (blue) curve represents an extremal charged black hole with $r_m=2~M$. Thus, increasing values of charge {moves} $r_m$ towards the Schwarzschild radius $r_s$. {On the other hand, if the electric charge is replaced by the tidal (or Weyl) charge in a braneworld scenario, the behavior of the potential is reversed: as this charge increases, the potential barrier lowers. This suggests that the influence of the extra dimension facilitates the penetration of photons into the black hole's interior.} 

After some algebra, Equation (\ref{eq6}) can be written in the following way:
\begin{equation}
\left(\frac{dr}{d\varphi}\right)=\pm r\sqrt{\frac{r^2}{b^2}-\frac{1}{A(r)}},
    \label{eq9}
\end{equation}
where $b \equiv \frac{L}{E}$ is the impact parameter of the photon. By integrating Equation (\ref{eq9}), the solution, $r(\varphi)$, describes the orbital motion of photons in terms of the constants of motion.   
In the closest approach, denoted by $r_0$, that satisfies $r_0 \geq r_m$ and $\frac{dr}{d\varphi}\left.\right|_{r=r_0}=0$, Equation (\ref{eq9}) gives the relation

\vspace{-6pt}
\begin{equation}
b=r_0\sqrt{ A(r_0)}.
    \label{eq10}
\end{equation}
\vspace{-24pt}
\begin{figure}[ht]
\centering
\includegraphics[width=\linewidth]{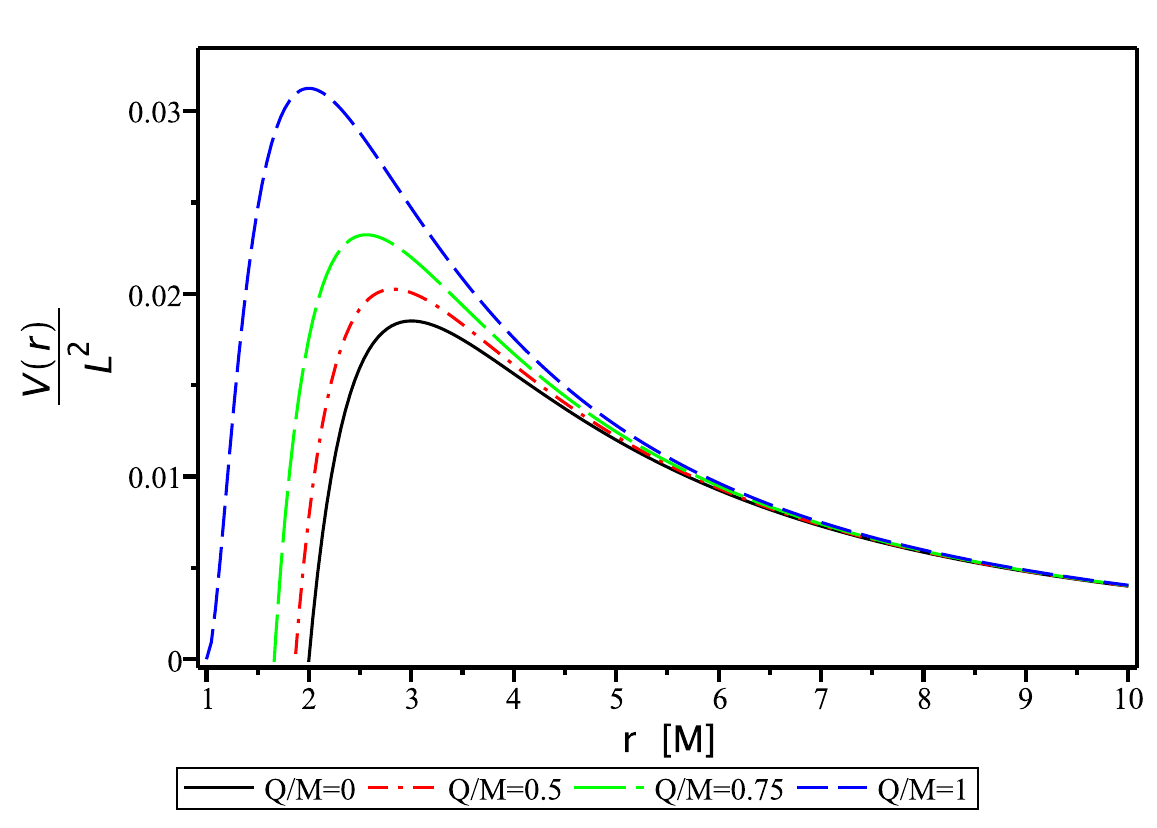}
\caption{Plots of the {effective}  potential as the function of $r$ in units of $M$ for four  fixed
values of $Q/M$. The solid line corresponds to the uncharged black hole. The charge of the black hole acts to reduce the values of $r_m$, as can be seen in the figure above.}
\label{fig1}
\end{figure}

Another useful expression related to $r_0$ is the critical impact parameter $b_c$, defined as
\begin{equation}
    b_c(r_m)\equiv \lim_{r_0\rightarrow r_m}{b(r_0)}.
    \label{eq11}
\end{equation}

{Indeed,}  the parameter $b_c$ is directly associated with the strong deflection limit since, at this limit, $r_m$ approaches $r_0$. With these quantities, the deflection angle $\alpha$  of the photon trajectory can be written in the following way \cite{naoki1}:
\begin{equation}
 \alpha=I(r_0) - \pi
 \label{eq12}
\end{equation}
where the integral $I$ is defined as twice the function $\varphi (r)$ obtained of Equation (\ref{eq9}), i.e., 
\begin{equation}
   I \equiv 2\int_{r_0}^{\infty}\frac{dr}{ r\sqrt{\frac{r^2}{b^2}-\frac{1}{A(r)}}}. 
    \label{eq13}
\end{equation}

{Equation} (\ref{eq13}), along with integral (\ref{eq12}), gives us the deflection angle in a  general way. Due to the type of trajectories that we are interested in, they will be solved by considering the strong gravitational limit where  $b\rightarrow b_C$. In this regime, the solution can be written in the following way (see \cite{naoki1} for more details):
\begin{equation}
\alpha (b)=-\bar{a}\log\left(\frac{b}{b_c}-1\right)+\bar{b}
    \label{eq14}
\end{equation}
up to higher-order terms in $(b-b_c)$, where the parameters $\bar{a}$ and $\bar{b}$ are given by

\begin{equation}
\bar{a}=\frac{r_m}{\sqrt{3Mr_m-Q^2}},
\label{eq15}  
\end{equation}
and 
\begin{equation}
\bar{b}=\bar{a}\log\left[\frac{8(3Mr_m-4Q^2)^3Z^2}{M^2r_m^2(Mr_m-Q^2)^2}\right]-\pi,
\label{eq16}  
\end{equation}
{where $Z=2\sqrt{Mr_m-Q^2}-\sqrt{3~Mr_m-4Q^2}$. In the uncharged case, corresponding to $Q=0$, Equations (\ref{eq15}) and (\ref{eq16}) reduce to the well-known results $\bar{a}=1$ and $\bar{b}= \log[216(7-4\sqrt{3})]=\pi$ respectively \cite{naoki1}.} 

 \section{Boomerang Photons}\label{sec4}
 Concerning the trajectories of the light around compact objects, a fundamental question arises: How can the light return to its emitter? As in the case of the geodesic of massive bodies, the light trajectory presents a rich structure of orbital shapes, depending on initial parameters. In this section, we explore some aspects of Equation (\ref{eq9}) associated with the orbit of photons that return to their emitter and find out the conditions for this type \mbox{of motion.} 

 \subsection*{Numerical Solution for the Observer Near the Black {Hole} }
 Based on the previous results, the orbits in which the light returns to  the emission point  can be determined. The motion is obtained by numerically solving Equation (\ref{eq9}) with its sign reversed as a turning point. The orbits of interest are generated by choosing an emission angle that satisfies $\beta>\beta_c$ {(the derivation of the expression for the emission angle $\beta$ can be found in Appendix \ref{appenB})}. In Figure \ref{fig2}, there are four orbits with different values of $r_i$ (emitter position) and $\beta$. If one considers a source in a planet placed at $r_i=10.5~M$, where $M$ is the black hole mass, the photons leaving the emitter to do a $\pi$ rotation about the charged black hole and then return to the planet. 
 
%
\begin{figure}[ht]
\centering
\includegraphics[width=\linewidth]{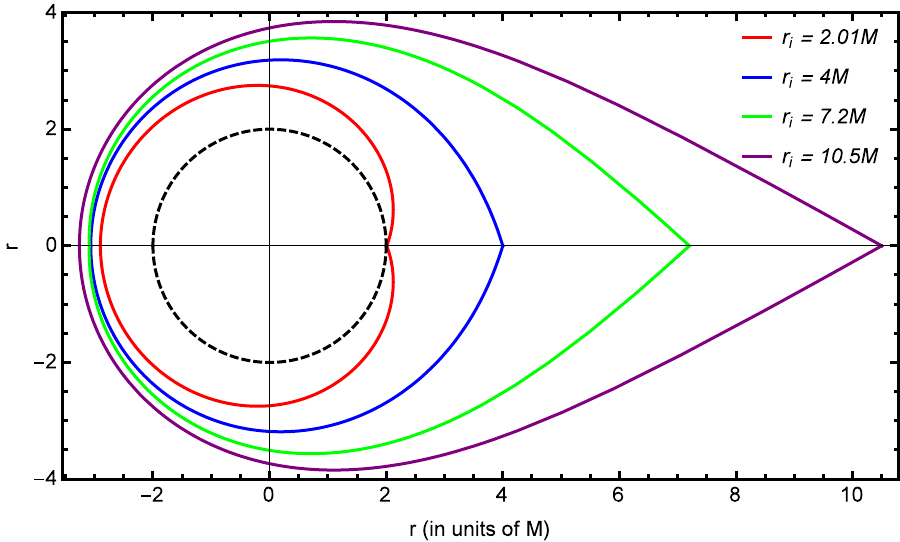}
\caption{Plots of the {trajectories}  of photons for four values of $r_i$ and $\beta$. The trajectory corresponding to photons emitted at $r_i=2.01$ with $\beta=0.1838$ {$rad$}  is inside the photon sphere and outside the horizon (dashed line). The other emission points, corresponding to $r_i>2$, are outside the photon sphere. For simplicity, the trajectories are obtained considering $Q=0$ in this figure. }
\label{fig2}
\end{figure}

\begin{figure}[ht]
\centering
\includegraphics[width=\linewidth]{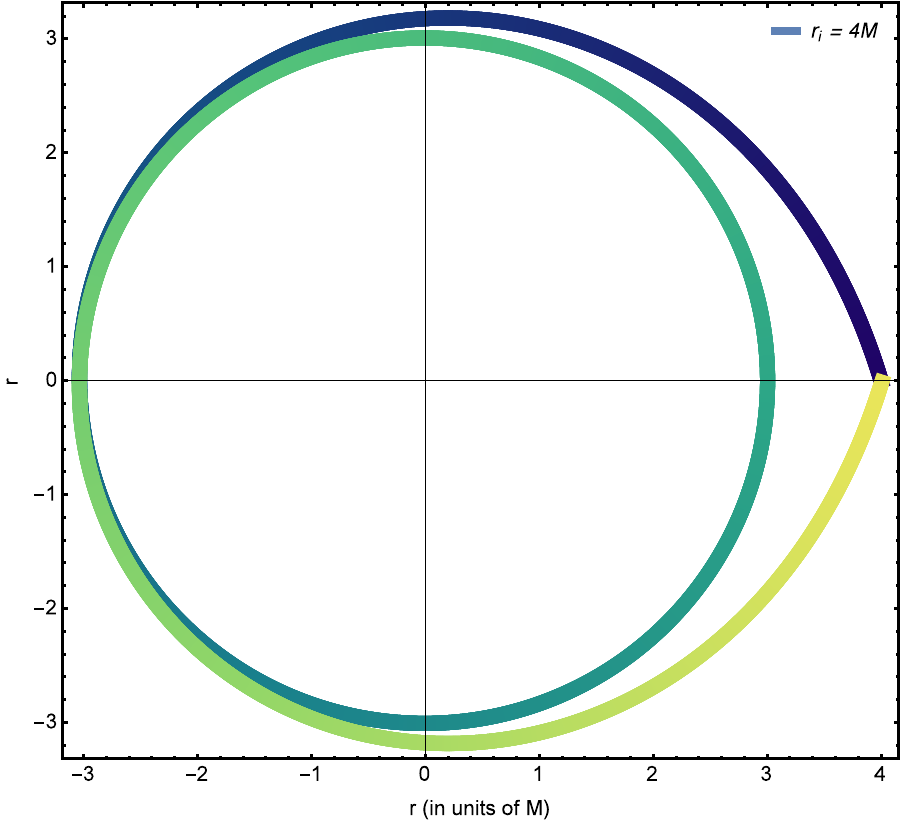}
\caption{Plot of the trajectory of photons emitted at $r_i=4.0$ with $\beta=1.164423$ $rad$. In this case, the emitted photons start in the blue region with such an angle that they will spiral and reach the turning point (in the green region), and then they arrive at point $r_i=4.0$ in the yellow region after two orbits around the black hole. }
\label{fig2b}
\end{figure}

 As shown in Table \ref{tab1}, the emission angle $\beta$ associated with this trajectory is {$\beta=0.466$ $rad$.}  Note that the dashed circle represents the event horizon of the black hole, where, for simplicity, we have considered the uncharged case. Small variations of $\beta$ can imply a different motion; for example, considering an emission with an angle of {$1.164423$ $rad$} at $r=4$, the photons made two orbits before returning to the emission point. This situation is illustrated in Figure \ref{fig2b}. 
 
\begin{table}[ht]
\caption{In this table are shown the parameters used in the numerical solution associated with the orbit of photons in Figure \ref{fig2}. Notice that the photons emitted at $r_i=2.01M$ are inside the photon sphere (with $Q = 0$) and near the horizon of an uncharged black hole.}
\label{tab1}
\begin{ruledtabular}
\begin{tabular}{c c c c}
$\boldsymbol{r_i/M}$ & $\boldsymbol{\beta}$ & $\boldsymbol{A(r)}$ & $\boldsymbol{b/M}$ \\
\hline
\midrule
2.0100 & 0.1838 & 201    & 5.2055 \\
4.0000 & 1.1664 & 2.0000 & 5.1997 \\
7.2000 & 0.6650 & 1.3846 & 5.1945 \\
10.500 & 0.4667 & 1.2353 & 5.2436 \\
\end{tabular}
\end{ruledtabular}
\vspace{-9pt}
\end{table}

%

The scheme in which returning photons are obtained by choosing an appropriate emission angle is more suitable when the emitter is close to the black hole. In a realistic scenario, considering our planet in the solar system, there are no known nearby black holes with the distances used in Figure \ref{fig2}. As we will see in the next sections, the study of the deflection of the light of the solar system bodies by black holes can be approached as a retrolensing problem. In this case, the black hole mass is an important parameter that can act to increase the magnitude of the retrolensing effect. 

{It is worth noting that although the addition of charge \( Q \) (or \( W \)) does not produce photon trajectories that differ significantly from the chargeless case, we can still qualitatively analyze the effect. Consider, for instance, Equation (\ref{eq22}) and the graph in Figure \ref{fig2}. If the electric charge is non-zero, the metric coefficient \( A(r) \) decreases, leading to an increase in angle \( \beta \). In contrast, when dealing with the tidal charge (or Weyl), which has the opposite sign to the electric charge, \( A(r) \) increases, resulting in a decrease in \( \beta \). In other words, we can interpret that, in the first case, the emitter needs to launch the photon at a slightly larger angle to receive it back, whereas in the second case, a slightly smaller angle is required for the same initial radius \( r_i \).
Thus, we have obtained a study showing the movement of photons close to black holes. This analysis demonstrates that it is possible to consider photons that return to the point of emission. In the next section, we will use the expertise gained here to study the dynamics of photons that are very far from the black holes.
}

\section{Black Holes as Gravitational Mirrors: A Possible Apparatus}\label{sec5}

\begin{figure}[ht]
\centering
\includegraphics[width=\linewidth]{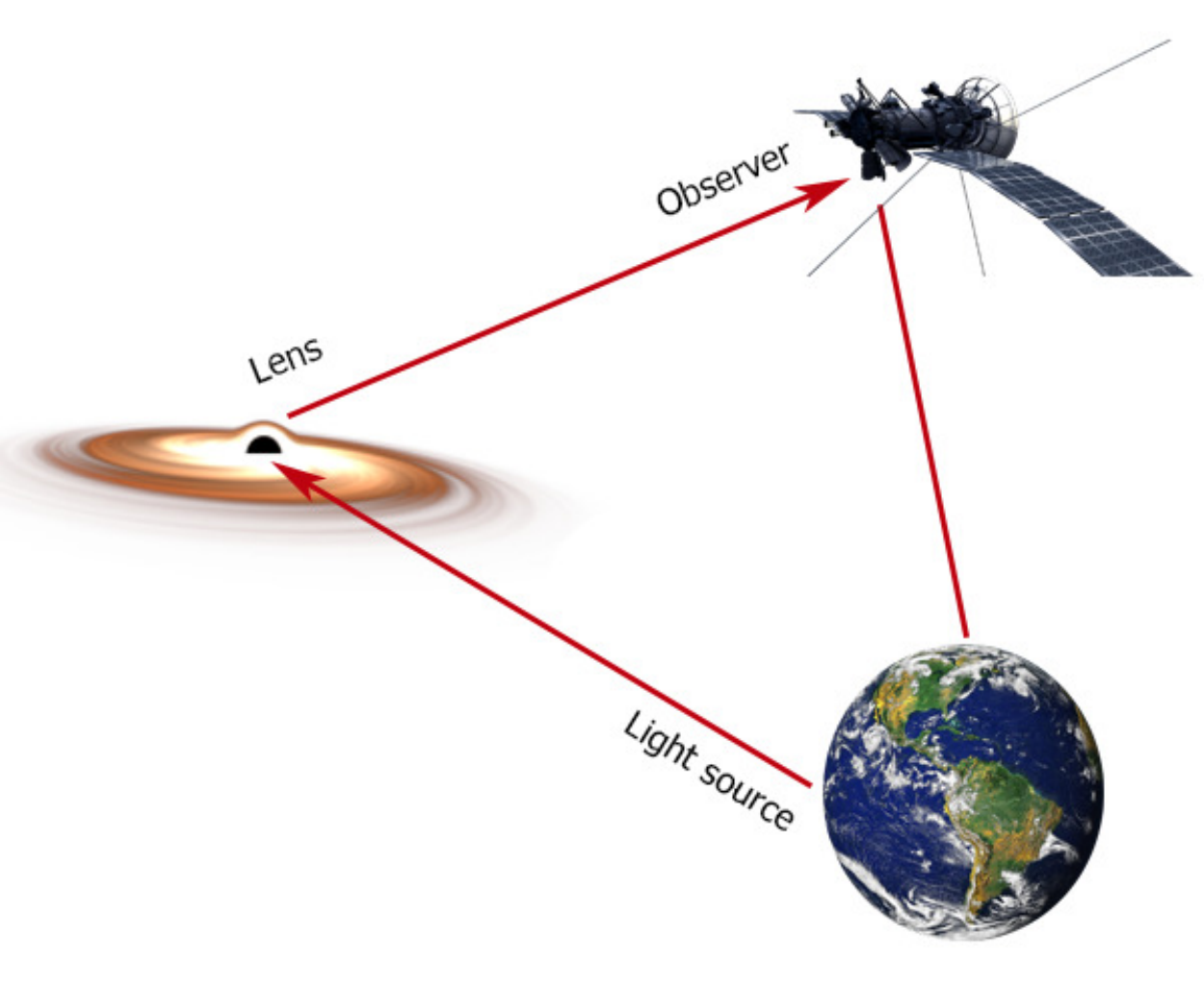}
\caption{The proposed scheme to address the lens geometry associated with the Earth as a source and an orbiting satellite as an observer. The emitted light from the Earth reaches the black hole, returning to the observer after receiving an amplification due to the strong gravitational field of the black hole.}
\label{fig3}
\end{figure}

In recent days, two geometries have been considered in the literature on gravitational lensing: the first one corresponds to a light source behind the black hole (standard geometry), while the
second one (retrolensing) presents a source in front of the black hole. In Figure \ref{fig3}, we illustrate the lensing configuration for a retrolensing geometry that we use in this work, where the Earth is represented as the light source for the retrolensing. The photons emitted reach the black hole, and then they return to the observer (satellite) near the Earth. Before we present the standard  retrolensing configuration, we discuss observing strategies based on the population of stars that probably collapse to form black holes locally.   
\vspace{-6pt}

As explained before, we will use a compact object with a strong gravitational field such that the light circulates around $n$ times before returning to the emission place; the equations associated with this type of motion indicate massive objects as lens candidates. Thus, {two categories} of black holes can be analyzed in our retrolensing study: distant supermassive black holes and nearby stellar black holes. The supermassive black hole in Sgr A* with a mass of $(4.1 \pm 0.014)\times 10^6 M_{\odot}$ \cite{retro26} is an example of a supermassive black hole that we analyze in the next section. The great mass of this object favors the amplification effect of the lens, but the big distance is a negative factor in this scenario. Another interesting aspect of this class of black holes is related to the age of the image observed. If the Earth is considered a light source in this retrolensing phenomenon, then we can see how Earth's light was {52,000}  years ago when Neanderthals lived on the Earth.
 
Despite the interesting implications of this scenario, this type of observation is very difficult with current technologies due to the distance from Sgr A* or another massive black hole. As a second possibility, we can consider nearby black holes. As discussed in \cite{retro14}, theoretical estimates give a result for the local density of stellar black holes {\it {circa} } $\sim 8\times 10^{-4}/pc^3$, while the mass of these objects lies in the range $5~M_{\odot} < M < 15~M_{\odot}$. The~earliest population of stars with masses in the range  $300 <M< 1000~M_{\odot}$ is another route to generate a retrolensing event since these stars probably collapse and form black holes with similar mass. 
At the moment, the nearest known black hole candidate  is the object HR6819 ($340\;pc$), but it has not been confirmed yet \cite{retro23}.

\begin{figure}[ht]
\centering
\includegraphics[width=\linewidth]{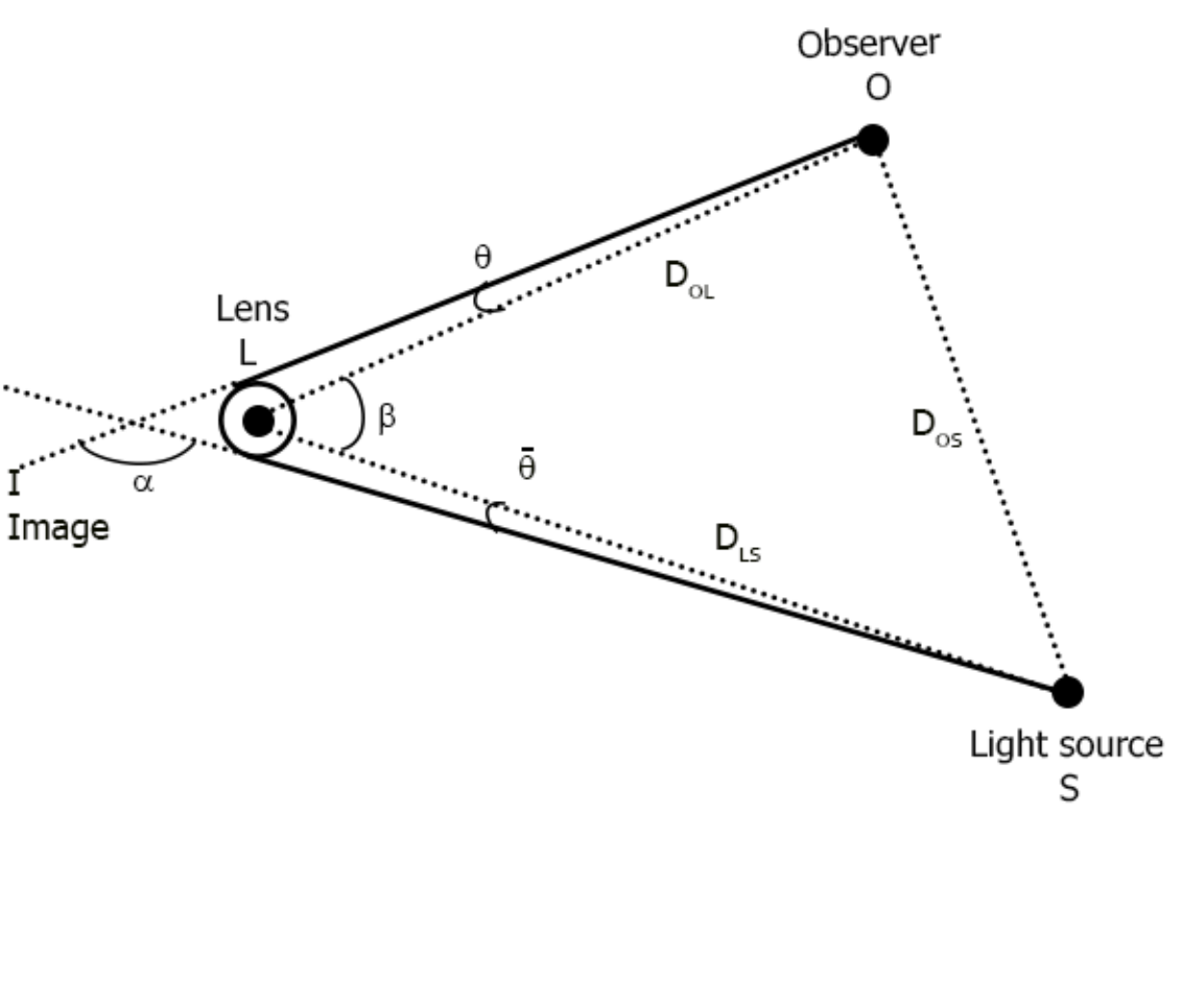}
\vspace{-38pt}
\caption{Retrolensing setup. The {source}  at S emits light rays to the lens L that turn around the black hole  several times before reaching the observer O.  The observer sees the light source as I, and from his perspective,  $\theta$ is the angle  of the image I. }
\label{fig4}
\end{figure}

In this section, we discuss the framework used to deal with retrolensing in the Reissner–Nordström space–time following Ref. \cite{naoki1}{. Then}, we consider the application
of the obtained results to the original retrolensing problem proposed in this paper. To realize the idea illustrated in Figure \ref{fig3}, we consider the retrolensing scheme  in   Figure \ref{fig4}. As we can see, the parameters $D_{OS}$, $D_{OL}$, and $D_{LS}$  are the distances between the observer and the Earth, between the observer and the black hole, and between the black hole and the Earth,  respectively. In addition, the angle $\theta$ is the angular position of the image with respect to the position of the black hole, $\beta$  is the angle formed between $D_{OL}$ and $D_{LS}$, and $\bar{\theta}$ is the angle between $D_{LS}$ and the light ray (see Figure \ref{fig4}). These angles are related by the Ohanian equation in the form \cite{retro24,retro25} 
 \begin{equation}
     \beta=\pi - \alpha(\theta)+\theta+\bar{\theta}.
      \label{eq23}
 \end{equation}
\vspace{-24pt}
 By assuming that the lens, the observer, and the source are almost aligned and {$b_c \ll D_{OL}$,} we can use the relations $\beta \sim 0$ and $D_{LS}=D_{OL}+D_{OS}$. It follows from  (\ref{eq23}) under the assumption of a strong deflection limit, neglecting small terms in $\theta$ and $\bar{\theta}$, the positive solution \cite{naoki1}
 
 \vspace{-4pt}
 \begin{equation}
     \theta_{+}(\beta)=\theta_{m}\left[1+\exp\left(\frac{\bar{b}-\pi+\beta}{\bar{a}}\right)\right]
      \label{eq24}
 \end{equation}
 where $\theta_{m}=b_{c}/D_{OL}$. The magnification associated to the Equation (\ref{eq24}) is given by \cite{naoki1}
 \begin{equation}
     \mu_{+}(\beta)=-\frac{D_{OS}^{2}}{D_{LS}^{2}}s(\beta)\theta_{+}\frac{d\theta_{+}}{d\beta}
      \label{eq25}
 \end{equation}
 where 
 
 \begin{align}
     s(\beta)&=\frac{2}{\pi \beta_{s}}\Big[\pi(\beta_{S}-\beta) \Big.\\ \nonumber
     &\left. +\int_{-\beta+\beta_{S}}^{\beta+\beta_{S}}\arccos{\frac{\beta^2+\beta'^2-\beta_{S}^2}{2\beta\beta'}}d\beta'  \right]
      \label{eq26}
 \end{align}
 for the case $\beta \leq \beta_S $, and 
 \begin{equation}
 s(\beta)=\frac{2}{\pi \beta_{s}}\int_{\beta-\beta_{S}}^{\beta+\beta_{S}}\arccos{\frac{\beta^2+\beta'^2-\beta_{S}^2}{2\beta\beta'}}d\beta'
      \label{eq27}
 \end{equation}
 for the case $\beta_S \leq \beta $, where $\beta_{S} \equiv R_{S}/D_{LS}$, with $R_{S}$ being the radius of the source. It is considered in Equation  (\ref{eq27})  that the origin of the coordinates is on the intersection point between the source plane and the axis $\beta =0$. The negative solutions for $\theta$ and $\mu$ can be written by observing the relations
 \begin{equation}
 \theta_{-}(\beta) \sim - \theta_{+}(\beta),\;\;\mu_{-}(\beta) \sim - \mu_{+}(\beta).
      \label{eq28}
 \end{equation}
 
 {In this} please confirm if correct. Following highlights are same. Authors:  Yes, this is correct.
 way, the total magnification $\mu(\beta)=|\mu_{+}(\beta)|+|\mu_{-}(\beta)|$ is obtained as the combination of these relations, the result is
 \begin{equation}
 \mu(\beta)=2\frac{D_{OS}^{2}}{D_{LS}^{2}}\frac{\theta_{m}^{2}e^{(\bar{b}-\pi)/\bar{a}}\left[1+e^{(\bar{b}-\pi)/\bar{a}}\right]}{\bar{a}}|s(\beta)|.
     \label{eq29} 
 \end{equation}
 
 {These} 
 results have been applied in the study of retrolensing problems involving the Sun as the source. For example, by considering a black hole of $M=10~M_{\odot}$ at distance {$D_{OL}=0.01 pc$} 
 in the case of perfect alignment, we obtain the maximum amplification with a magnitude of $m \sim 26$, an observable value. In the next section, we apply these ideas, considering that the Earth moves on the source plane with the orbital velocity.

\subsection*{{Retrolensing} } 
\begin{figure}[ht]
\centering
\includegraphics[width=\linewidth]{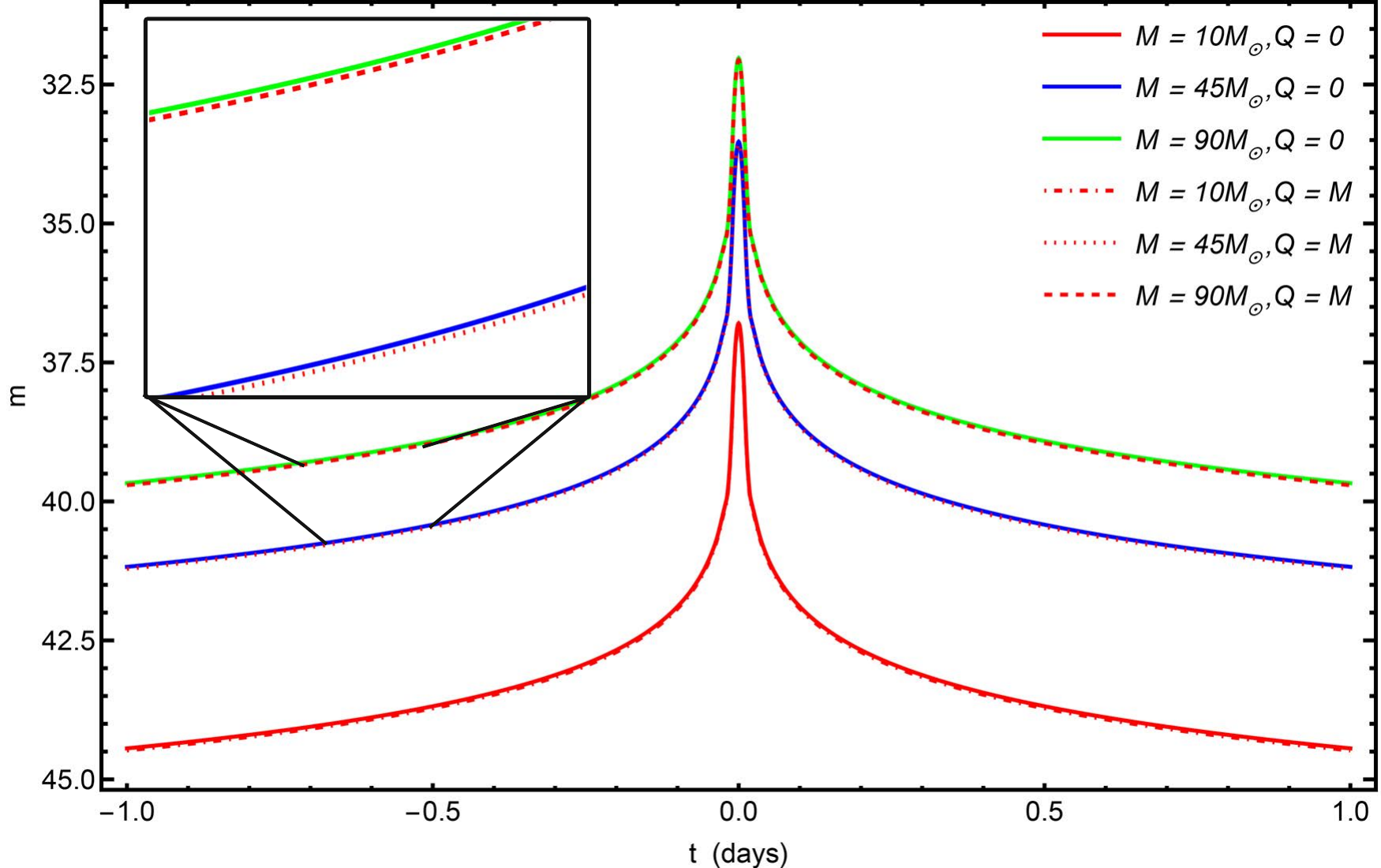}
\caption{Light curves by non-charged and charged  black holes for different masses at a distance of 0.001 pc. The closest separation considered is $\beta=0$. The green, blue, and red curves represent the magnification of black holes with masses $90~M_{\odot}$, $45~M_{\odot}$, and $10~M_{\odot}$, respectively.}
\label{fig5}
\end{figure}

\begin{figure}[ht]
\centering
\includegraphics[width=\linewidth]{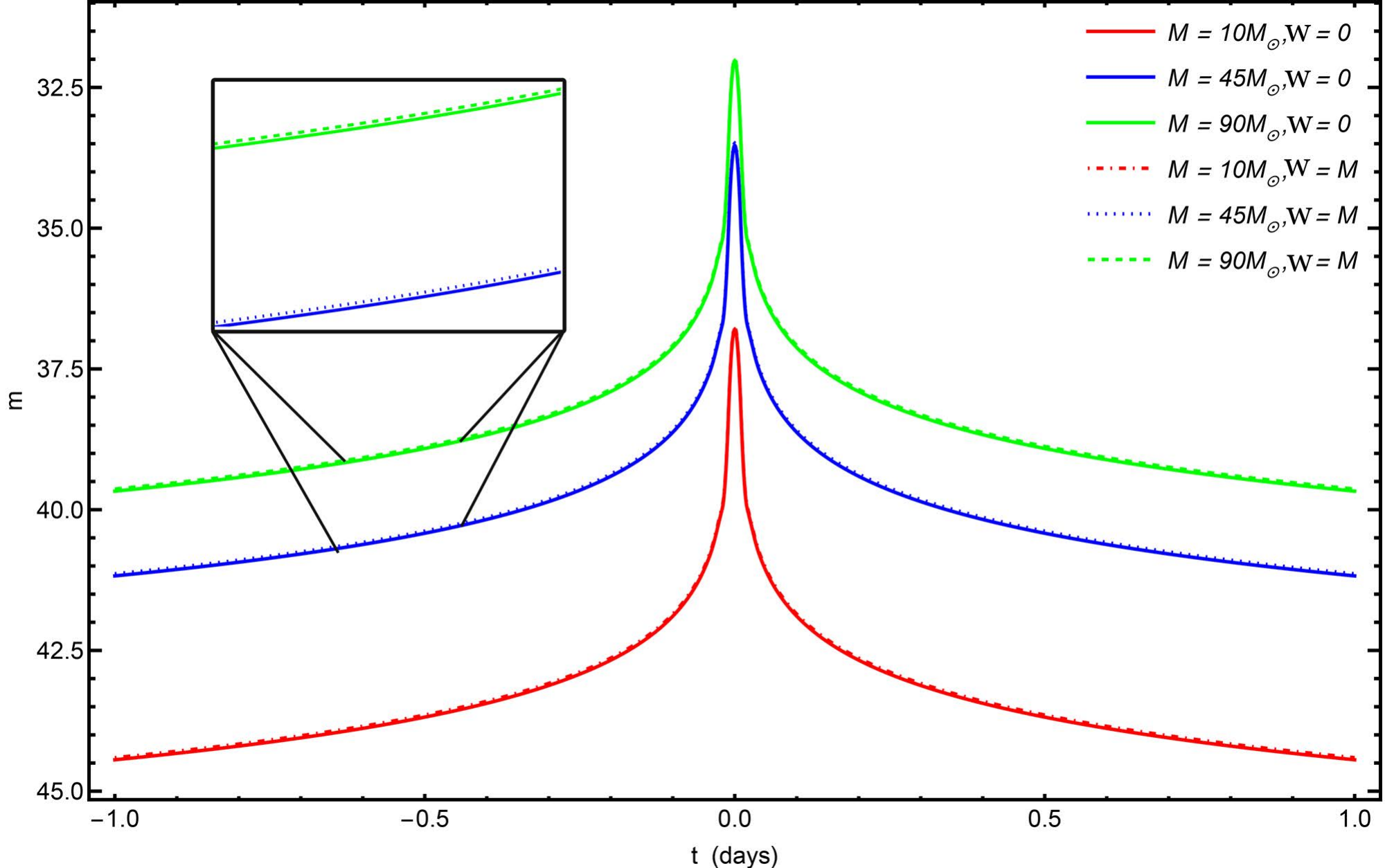}
\caption{Light curves by a black hole with the Weyl tidal charge for different masses at a distance of 0.001 pc. The closest separation considered is $\beta=0$. The green, blue, and red curves represent the magnification of black holes with masses $90M_{\odot}$, $45M_{\odot}$, and $10M_{\odot}$, respectively.}
\label{fig5b}
\end{figure}

Now, we investigate the retrolensing light curves by nearby black holes as proposed by \cite{retro14} with masses in the range $(10-90)M_{\odot}$ considering the scheme proposed in {Figures \ref{fig3} and \ref{fig4}.} By inspection of Equation (\ref{eq29}), it {can be seen} that the parameters $D_{LS}$ and $M$ have played an important role in the final value for the magnitude of the images.  In~this first example, {we consider} an observer placed at $D_{OS}=380,000$  {km} of distance from the Earth (source). Figure \ref{fig5} shows the magnitude of the image considering three different masses {and  two different charges.} The best value, $m \sim 32$ for magnitude, is obtained with the $90~M_{\odot}$ black hole at a distance of 0.001 pc {with $Q=0$. As we can see, the curves associated with the extremal charge $Q = M$ have a maximum value for $m$ lower than the uncharged case. For comparison, we consider in Figure \ref{fig5b} the magnitude of the images for the case where we replace the electric charge with the Weyl tidal charge. In this case, we have an opposite behavior; the charge increases to the maximum value of $m$.}  
 
 \vspace{4pt}

\begin{figure}[ht]
\centering
\includegraphics[width=\linewidth]{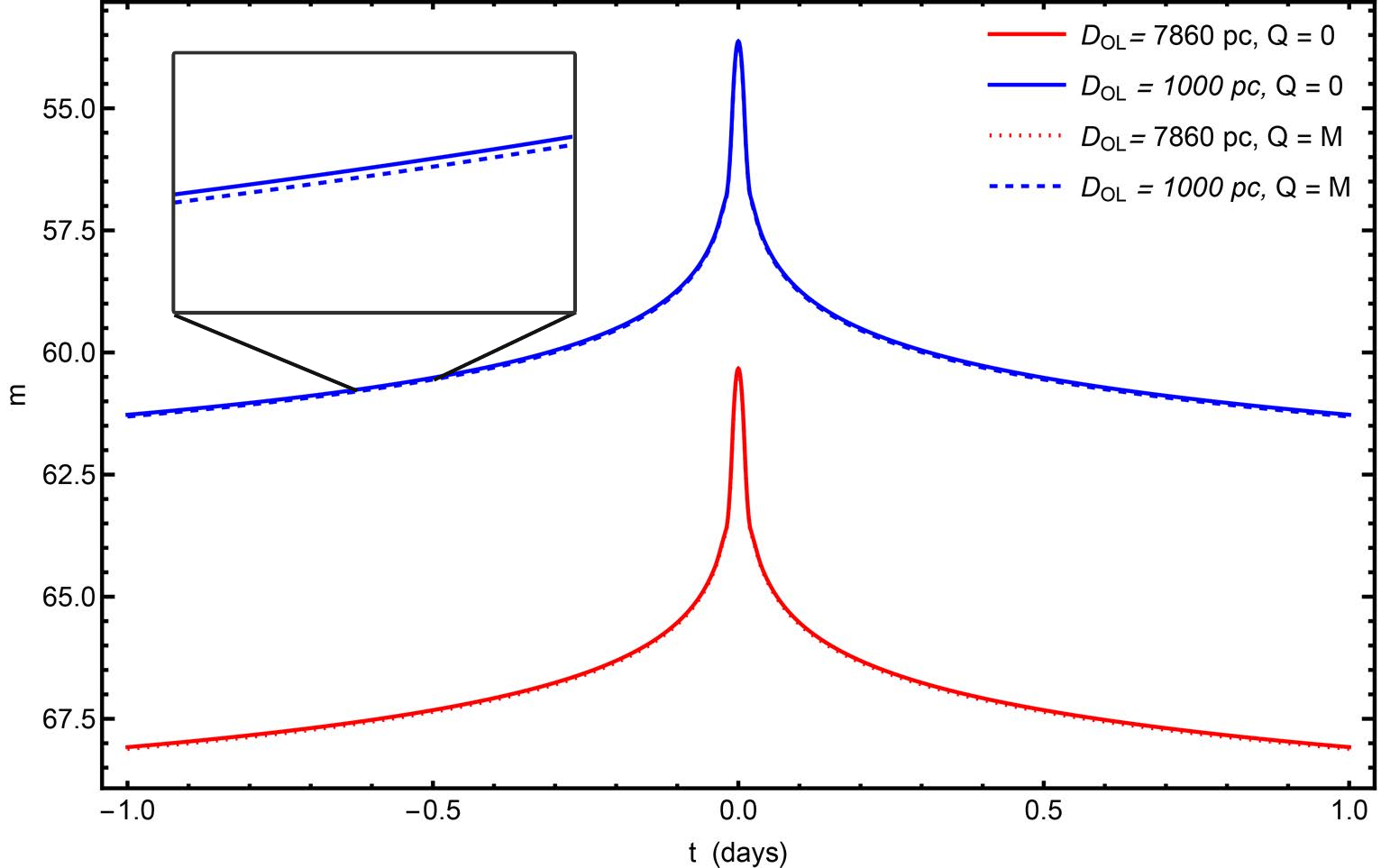}
\caption{Light curves by a supermassive  black hole for different distances with $4.02\times 10^6~M_{\odot}$ and two different charges. The closest separation considered is $\beta=0$. Red and blue curves represent the magnification of black holes at $7860$ pc and $1000$ pc,  respectively.}
\label{fig6}
\end{figure}
\begin{figure}[ht]
\centering
\includegraphics[width=\linewidth]{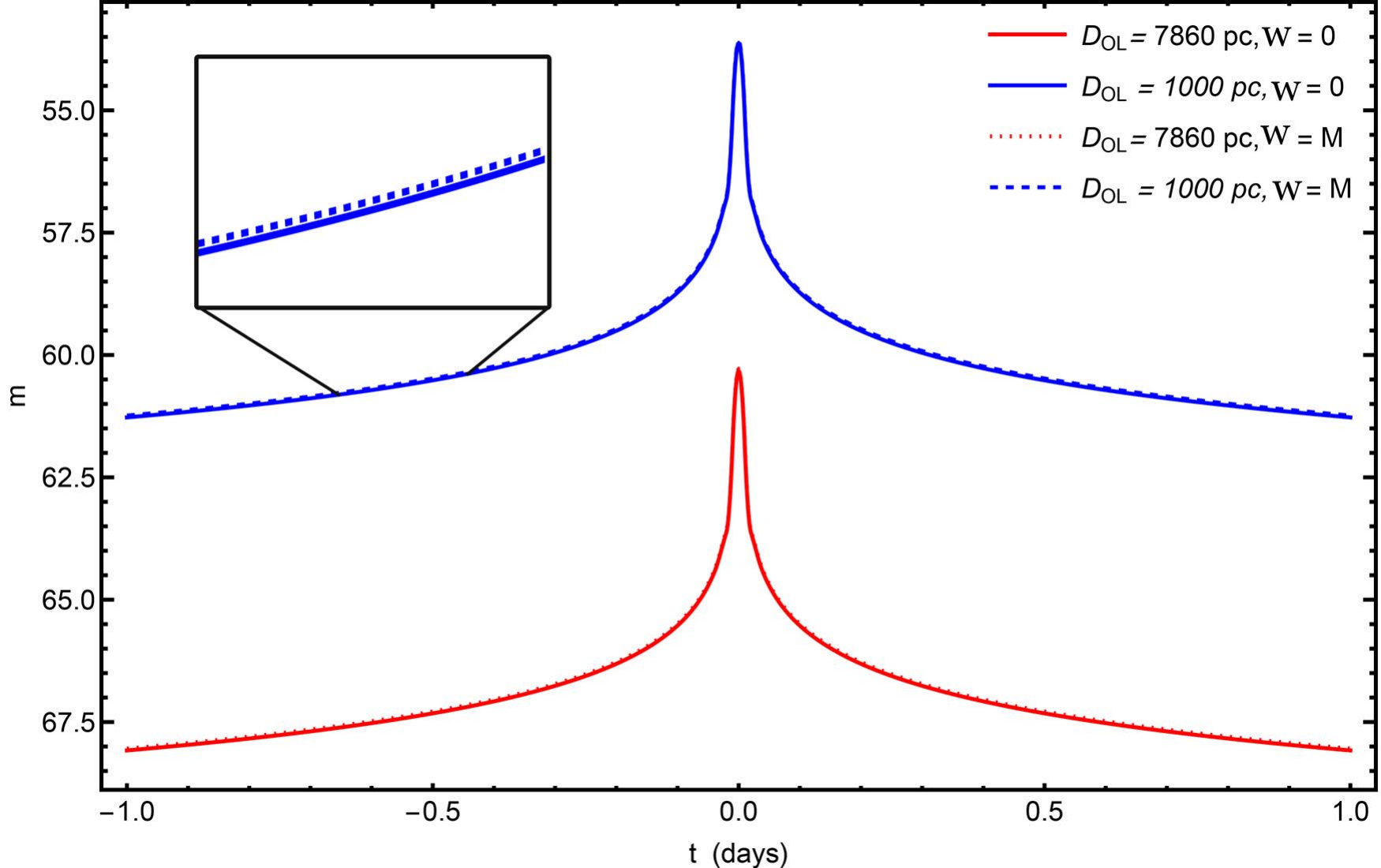}
\caption{Light curves by a supermassive  black hole with the Weyl tidal charge for different distances with $4.02\times 10^6 M_{\odot}$ and two different charges. The closest separation considered is $\beta=0$. Red and blue curves represent the magnification of black holes at $7860$ pc and $1000$ pc,  respectively.}
\label{fig6b}
\end{figure}

At a distance of $7860$ pc, the supermassive black hole in  Sgr A* is possibly the central black hole of the Milky Way. Due to the value of its mass, the study of gravitational retrolensing in space–time provides a convenient tool to explore the strong deflection limit. {Figure \ref{fig6}} shows the magnitude of the light curves considering the black hole in Sgr A* (red line) and a black hole with the same mass but at 1000 pc (blue line) {for two different charges. In this case, the charge of the black hole reduces the value of the maximum magnitude. Figure \ref{fig6b} shows the magnitude of the images for the case where we replace the electric charge with the Weyl tidal charge. In this case, the charge increases to the maximum value of $m$.} As we can see, the magnitudes of the retrolensing light curves are very difficult to detect with current instruments in this case, even considering a supermassive black hole at 1000 pc.

\vspace{-2pt}

\vspace{-6pt}

\section{Conclusions}\label{sec6}
Considering the trajectories of photons in the spherically symmetric space–time, we conclude that photons can return  to  their  emitter, the so-called  boomerang photons. This is the effect of the strong gravitational field  that imposes an intense influence on the  geodesic  equation  for  the  bending  of  light. Based on this result, we suggest a particular configuration where the emitter is the Earth itself.  We have shown {that} these trajectories can be obtained in terms of the constants of the motion (Figures \ref{fig2} and \ref{fig2b}). {We also saw that when the electric (tidal) charge is included, the emission angle increases (or decreases) because the height of the effective potential for the photon becomes higher (or lower).} Depending on the emission angle, i.e., the angle between the propagation direction of the photon and the radial direction, it can rotate $N$
times around the black hole before returning to the emission point. This type of motion can be used in principle as a tool to determine the mass of a black hole as well as the distance of the observer to this object \cite{muller}. 

Based on parameters used in the numerical solution associated with the orbit of photons in the figures, we can see that the photons emitted at $r_i= 2.01~M$ are inside the  photon sphere  and  near the  horizon  of  an uncharged {black hole. The highest value of the initial radius that we used} was $r_i= 10.5~M$. In a more realistic scenario, considering the Earth as a source, the setup used in retrolensing systems is the most adequate  to address the study of returning light rays, where we can use distant supermassive black holes as lenses. Different from the standard gravitational lensing, retrolensing permits the light rays emitted by Earth to be reflected from the light sphere of the black hole.

Although the probability is small, a close approach of a stellar-mass black hole, besides potentially catastrophic consequences (effects on the orbital stability), can  offer an observational verification of the Earth as the source of a retrolensing event. Figure \ref{fig5} shows that this retrolensing event can be seen with the apparent magnitude $m=32$. In a safer scenario by considering the Sgr A* black hole at 7860 pc, Figure \ref{fig6} shows a magnitude  $m=61$. In the same figure, a hypothetical black hole with the same mass at 1000 pc has a magnitude around $m\sim 53$.

{When electric ($Q$) or Weyl tidal ($W$) charges are taken into account, the resulting change in magnitude relative to the charge-free case remains negligible. In particular, our analysis indicates that the presence of a tidal charge leads to a slight increase in the luminosity of the recovered image (consider Figure~\ref{fig6b}). In this case, by adopting the upper bound $W^2 \lesssim 0.0004\, M^2$, as reported in Ref.~\cite{REF1}, the corresponding magnitude difference is approximately $\Delta m \sim 0.0001$. This value is about one order of magnitude below the current threshold of astronomical photometric sensitivity~\cite{REF2}.
}

In future work, it is possible to consider space–time solutions with different symmetries where the cosmological constant and/or rotation of the matter are taken into account. 

The observational possibilities for retrolensing events are not restricted to charged static black holes or the aforementioned extensions. Interesting scenarios can be studied in modified theories of gravity \cite{retro28,santos8,retro27}, where new solutions arise, generalizing the usual geometries associated with wormholes and black holes. In this way, the observation of a retrolensing event, besides the confirmation of GR in a strong-field regime, can be used to constrain modified gravity theories in this field.

\appendix

\section[\appendixname~\thesection]{Lower Bound for Stable Orbits} \label{appenA}

 The classical motion of a test particle, in general space–time, can be analyzed considering a Lagrangian in the form $\mathcal{L}=(1/2)g_{\mu\nu}\dot{x}^{\mu}\dot{x}^{\nu}$, where the dot denotes the differentiation
with respect to the affine parameter. In this way, considering Equation (\ref{eq1}) and assuming $\theta=\pi/2$, as usual, one obtains 
\begin{equation}
    2\mathcal{L}=-B(r)\dot{t}^2+A(r)\dot{r}^2+r^2\dot{\varphi}^2.
    \label{eq4}
\end{equation}

{Due} to the spherical symmetry of the system under consideration, the translational Killing vector   $\partial_t=t^{\mu}\partial_{\mu}$ {with $t^{\mu}=(1,0,0,0)$} and the axial Killing vector $\partial_{\varphi}=\varphi^{\mu}\partial_{\mu}$ {with {$\varphi^{\mu}=(0,0,0,1)$}} yield the constants of the motion 
\begin{equation}
 E=-g_{\mu\nu}t^{\mu}\dot{x}^{\nu}=-B(r)\dot{t}, 
    \label{eq5}
\end{equation}
and
\begin{equation}
L=g_{\mu\nu}\varphi^{\mu}\dot{x}^{\nu}=r^{2}\dot{\varphi},
    \label{eq6.1}
\end{equation}
 respectively. In a general way, $E$ can be interpreted as the total energy per unit rest mass of the particle in the case of a timelike geodesic. In the case of null geodesics, the product $\hbar E$ is the total energy of a photon. Similarly, $\hbar L$ represents the total angular momentum of a photon in the case of a null geodesic. By considering the constants of motion (\ref{eq5}) and (\ref{eq6.1}) along the null geodesic condition  $\mathcal{L}=(1/2)g_{\mu\nu}\dot{x}^{\mu}\dot{x}^{\nu}= 0$, Equation (\ref{eq4}) may be written in the form
 \begin{equation}
 \frac{\dot{r}^2}{2}+V(r)=\frac{1}{2}\frac{E^2}{A(r)B(r)},
     \label{eq6}
 \end{equation}
 where 
 \begin{equation}
     V(r)=\frac{1}{2}\frac{L^2}{A(r)r^2}.
     \label{eq7}
 \end{equation}

{This} is the potential associated with the photon trajectory. An important region of this motion is around the extremum of $V(r)$, the point $r_m$, where  the differentiation with respect to the radial coordinate vanishes. An expression for this region can be written as

\begin{equation}
r_m=\frac{3M}{2}+\frac{\sqrt{9M^2-8Q^2}}{2}.
    \label{eq8}
\end{equation}
This region is the lower bound for stable orbits in this space–time.

\section[\appendixname~\thesection]{Emission Angle} \label{appenB}
By considering an observer at rest, the line element (\ref{eq1}) provides the relation 
\begin{equation}
ds^2=-d\tau^2=-B(r)dt^2.
    \label{eq17}
\end{equation}

{In} 
 addition, the null geodesic condition ($ds^2=0$) permits us to write the line element in the form 
\begin{equation}
-B(r)dt^2=A(r)dr^2+r^2d\theta^2+r^2\sin\theta^2d\varphi^2
    \label{eq18}
\end{equation}

{Finally,} identifying Equations (\ref{eq17}) and (\ref{eq18}) considering the equatorial motion, we get the result
\begin{equation}
d\tau^2=A(r)dr^2+r^2d\varphi^2,
    \label{eq19}
\end{equation}
that, after some algebra, can be recast as
\begin{equation}
\left(\sqrt{A(r)}\frac{dr}{d\tau}\right)^2 + \left(r\frac{d\varphi}{d\tau}\right)^2
=1^2,    \label{eq20}
\end{equation}

{The} two terms on the left-hand side of Equation (\ref{eq20}) can be identified with the radial and angular components of the speed of light $(c=1)$. In this case, the components $V_r$ and $V_\varphi$ are related to the angle between the propagation direction and radial direction  in the following way:
\begin{equation}
\frac{V_r}{V_\varphi}=\tan\beta=\frac{r}{\sqrt{A(r)}}\frac{d\varphi}{dr}, 
   \label{eq21}
\end{equation}
where the term $d\varphi/dr$ can be replaced by the right-hand side term of Equation (\ref{eq9}), the result is
\begin{equation}
\tan\beta=\pm\left( \sqrt{\frac{r^2}{b^2A(r)^{-1}}-1}\right)^{-1}. 
   \label{eq22}
\end{equation}

{In this} way, an emitter at rest at $r$ measures an angle $\beta$  between the propagation direction of the photon and the radial direction that determines if the photons are captured or returned to  the emitter. The critical angle $\beta_c$, defined as the angle where the photon is captured in the photon circle, is a good parameter to analyze this type of motion. It acts as a threshold separating orbits where the photons are captured by the black hole, escape to infinity, or are captured in the photon sphere.

\acknowledgments
This research was funded by FAPESC under Grant No. 735/2024 (L.C.N.S.).

\bibliography{referencias_unificadas} %

\end{document}